\documentclass[prb,twocolumn,showpacs]{revtex4}
\usepackage{amsmath}
\usepackage{graphicx}

\newcommand{\be}{\begin{equation}}
\newcommand{\ee}{\end{equation}}
\newcommand{\eps}{\varepsilon}
\newcommand{\mb}[1]{\mathbf{#1}}

\newcommand{\nn}{\nonumber}
\newcommand{\ce}{CePt$_3$Si}

\newcommand{\Journal}[4]{#1 \textbf{#2}, #3 (#4)}
\newcommand{\PRev}{Phys. Rev.}

\begin{document}

\title{Order parameter in superconductors with non-degenerate bands}

\author{I. A. Sergienko}
\email[Electronic address: ]{sergienko@ornl.gov}
\altaffiliation[Present address: ]{Condensed Matter Sciences Division, Oak 
Ridge National Laboratory, Oak Ridge, Tennessee 37831, USA.}
\author{S. H. Curnoe}
\affiliation{Department of Physics and Physical Oceanography,
Memorial University of Newfoundland, St.\ John's, Newfoundland \& Labrador 
A1B 3X7, Canada}

\begin{abstract}
In noncentrosymmetric metals, the spin degeneracy of the
electronic bands is
lifted by spin-orbit coupling.
We consider general symmetry properties of the pairing function
$\Delta(\mb k)$ in noncentrosymmetric superconductors with 
spin-orbit coupling (NSC), including \ce, UIr and Cd$_2$Re$_2$O$_7$.
We find that $\Delta(\mb k) = \chi(\mb k) t(\mb k)$, where $\chi(\mb k)$ is
an even function which transforms according to the irreducible representations 
of the crystallographic point group and $t(\mb k)$ is a model dependent phase 
factor. We consider tunneling between a NSC and a conventional superconductor.
It is found that, in terms of thermodynamical properties as well as the 
Josephson effect, the state of NSC resembles a singlet superconducting state 
with gap function $\chi(\mb k)$. 
\end{abstract}


\maketitle

\section{Introduction}

A better understanding of unconventional superconductivity has been sought
ever
since the pioneering discovery of heavy fermion superconductivity in 
CeCu$_2$Si$_2$.\cite{Steglich79} Soon after, 
the effect of spin-orbit coupling (SOC)
was recognized to play an important
role in the superconducting (SC) properties of heavy fermion materials, 
because it breaks spin rotation symmetry.\cite{Volovik84,Anderson84}
If there is no spin-orbit coupling, {\em or} if
the crystal has space inversion and time reversal symmetry, 
the electronic bands are doubly degenerate everywhere in the Brillouin 
zone.\cite{Jones75} Then, the superconducting states can be 
classified as spin-singlet (even parity) and spin-triplet (odd 
parity), where the spin is replaced by pseudospin
when SOC is present. \cite{Volovik84,Volovik85,Ueda85,Sigrist91}
However, if inversion symmetry is absent {\em and} SOC is present, the 
degeneracy of electronic band states 
is removed at all points in the Brillouin zone except at some highly symmetric 
positions, while time reversal symmetry ensures that states with opposite 
momenta are degenerate.\cite{Jones75}

The recent interest in noncentrosymmetric superconductors
was stimulated by the discovery of superconductivity in \ce\ and 
UIr.\cite{Bauer04,Akazawa04} SOC is normally strong in 
cerium and uranium compounds since the atoms are heavy.
A recent band structure calculation of \ce\ found that the 
bands, which would be degenerate if SOC was absent, are split by 
50-200~meV for the states close to Fermi level.\cite{Samokhin04} 
This splitting energy is a 
factor of more than a thousand on the characteristic energy scale 
$k_BT_c$. 
Therefore, the one-band theory, which applies to
the spin-degenerate case, should be reformulated.\cite{Saxena04}

Actually, several superconducting materials in which inversion
symmetry is absent have been known for some time.
Sesquicarbide materials, with chemical formula
R$_2$C$_{3-y}$ (R = La or Y, which can be partially substituted by a number of
elements) have $T_c$'s of up to 18 K, and space group symmetry 
$I\bar 43d$.\cite{Krupka69,Giorgi70,Amano04}
This belongs to the
tetrahedral crystallographic class $T_d$ which has no 
symmetry centre. However, in sesquicarbides SOC does not seem to be important
for conduction electrons, {\em i.e.} the bands are still spin 
degenerate.\cite{Singh04,NoteTh}

In 2001, superconductivity with $T_c=1$ K was reported in 
Cd$_2$Re$_2$O$_7$.\cite{Hanawa01,Sakai01}
At room temperature, this material
has the ideal pyrochlore structure, which includes
a centre of
symmetry. However, superconductivity occurs on the background of a 
noncentrosymmetric tetragonal crystal field after a series of structural 
phase transitions.\cite{Yamaura02,Sergienko03,Sergienko04a} 
In addition, SOC dramatically affects the electronic band structure, as shown 
by first-principles calculations for the pyrochlore phase.\cite{Harima02,
Singh02} Following
further calculations,\cite{Eguchi02} it was predicted that if the centre of 
symmetry is removed from the structure, the spin-orbit splitting of the bands 
reaches  68 meV $\approx 700\, k_B T_c$. This result was indirectly 
verified by a photoemission study, which found that the 
energy spectrum is shifted noticeably towards higher binding energies at low 
temperatures as compared to room temperature.\cite{Eguchi02}
Therefore, Cd$_2$Re$_2$O$_7$ should be considered in the same category as
\ce\ and UIr. Moreover, Cd$_2$Re$_2$O$_7$ is a useful
test system for the theory
since its superconducting properties are well established.\cite{Lumsden02,
Vyaselev02} The theory must, therefore, include the possibility for a nodeless 
$s$-wave-like order parameter.\cite{SwaveNote} 
Another appealing feature of Cd$_2$Re$_2$O$_7$
is the relative simplicity of its normal metallic state: it neither orders
magnetically,\cite{Vyaselev02} nor demonstrates Kondo-like behaviour as
that found in \ce.\cite{Bauer04}

In this Article, we consider general symmetry properties of the superconducting
state in noncentrosymmetric crystals in which the degeneracy is lifted by 
strong SOC. As a first approximation,  we restrict 
consideration to a one-band model, neglecting possible pairing in other bands 
and inter-band interactions. Sec.~\ref{sec2} discusses the symmetry properties
of the superconducting order parameter. In Sec.~\ref{secJos}, we consider
Josephson tunneling between a conventional superconductor and a superconductor
with nondegenerate bands. Sec.~\ref{secLim} discusses the limiting case of 
small SOC, using the Rashba Hamiltonian as a specific example. 
In Sec.~\ref{secDisc}, we discuss the results of our approach in
relation to previous theoretical developments  and the 
applicability to real materials.

\section{\label{sec2}Superconducting order parameter}

Throughout this paper we use the weak coupling approach to treat the
pairing interaction. While this approach offers no insight about
the pairing 
mechanism, it has the advantage that many results can be obtained exactly.
Also, the symmetry properties of the SC state remain essentially the same as 
in strong coupling models.\cite{Sigrist91}

\subsection{Single-particle Hamiltonian}
In the condensed state it is normally sufficient to treat relativistic
effects by the Pauli approximation to Dirac's fully relativistic approach. 
We begin by briefly reviewing the properties of the single-particle 
Hamiltonian\cite{Jones75} 
\be\label{Ham1}
H_1=\frac{\mb p^2}{2m} + V(\mb r) + \frac{1}{4m^2c^2}(\nabla V(\mb r)
 \times \mb p \cdot \boldsymbol \sigma),
\ee 
where $\mb p = -i \nabla$ is the momentum operator in the 
coordinate representation, $V(\mb r)$ is the periodic potential of the crystal 
lattice, and $\boldsymbol \sigma = (\sigma_x, \sigma_y, \sigma_z)$ are the 
Pauli matrices. Here and in the following we use the unit system in which
$\hbar = k_B =1$. The space group of the crystal is defined as the set of 
operations $g$ (rotations, reflections and translations)
acting on the real space coordinates $\mb r$ which leave $V(\mb r)$ invariant,
\be
g: V(\mb r) = V(g \mb r) = V(\mb r).
\ee
Because of the last term in~(\ref{Ham1}), the transformation properties of 
$H_1$ are expressed as
\be
g: H_1(\mb r) = {\cal D}(g) H_1(g \mb r) {\cal D}(g)^{-1} = H_1(\mb r),
\ee
where ${\cal D}(g)$ is a $2 \times 2$ spin-1/2 rotation matrix. ${\cal D}(g)$ 
is defined as follows. If $g$ is a bare rotation about the axis $\mb n$ by an
angle 
$\phi$, then
\be
{\cal D}(g) = \cos \frac \phi 2 - i \mb n \cdot \boldsymbol \sigma \sin 
\frac \phi 2.
\ee
If $g$ includes inversion $I$, $g = I g'$, or a translation $\tau(\mb R)$, 
$g = \tau(\mb R) g'$, where $g'$ is a bare rotation, then 
$
{\cal D}(g) = {\cal D}(g'),
$
since inversion and translations do not change the spinor components.

The solutions $\Psi(\mb r)$ of the Shr\"odinger equation 
\be\label{eqShr}
H_1 \Psi(\mb r) = \eps \Psi(\mb r)
\ee
are two-component spinors which transform
according to
\be\label{transPg}
g: \Psi(\mb r) = {\cal D}(g) \Psi(g \mb r).
\ee
In nonmagnetic crystals $V(\mb r)$ does not depend on spin, and the Hamiltonian
(\ref{Ham1}) is also invariant under time reversal ${\cal K}$,
\be
{\cal K}:  H_1(\mb r) = (-i \sigma_y) H_1^*(\mb r) (-i \sigma_y)^{-1} = 
H_1(\mb r),
\ee
which uses the fact that spin reverses sign under
time reversal, expressed by  $\sigma_y\boldsymbol\sigma^*
\sigma_y=-\boldsymbol\sigma$.
Correspondingly,
\be\label{transPK}
{\cal K}: \Psi(\mb r) = (-i \sigma_y) \Psi^*(\mb r).
\ee
Further, if $\tau(\mb R)$ is a proper lattice translation,
\be
\tau(\mb R): H_1(\mb r) = H_1(\mb r + \mb R) = H_1(\mb r).
\ee
Therefore, as well as in the case of vanishing SOC, the solutions of the
Shr\"odinger equation~(\ref{eqShr}) are Bloch waves labeled by a 
quantum number quasimomentum $\mb k$,
\be\label{Psik}
\Psi_{\mb k}(\mb r) = U_{\mb k}(\mb r) e^{i \mb k \mb r},
\ee
where $U_{\mb k}(\mb r)$ is the periodic spinor
\be
U_{\mb k}(\mb r) =
\left(
\begin{array}{c}
u_{\mb k, \uparrow}(\mb r)\\
u_{\mb k, \downarrow}(\mb r)
\end{array}\right),
\qquad U_{\mb k}(\mb r+ \mb R) = U_{\mb k}(\mb r).
\ee
It follows from Eq.~(\ref{transPg}) that
\begin{eqnarray}
g: \Psi_{\mb k}(\mb r) &=& {\cal D}(g) 
U_{\mb k}(g \mb r) \, e^{i (\mb k, g \mb r)}\nn\\
&=&{\cal D}(g) 
U_{\mb k}(g \mb r) \, e^{i (g^{-1} \mb k, \mb r)}
\equiv \Psi^{(g)}_{g^{-1}\mb k}(\mb r).
\end{eqnarray}
Clearly, $\Psi^{(g)}_{g^{-1}\mb k}(\mb r)$ is a solution of (\ref{eqShr}),
corresponding to the quasimomentum $g^{-1}\mb k$. On the other hand, 
the same can be said about the function $\Psi_{g^{-1}\mb k}(\mb r)$ obtained 
from $\Psi_{\mb k}(\mb r)$ by the bare replacement 
$\mb k \rightarrow g^{-1}\mb k$. 
If the spin degeneracy of the bands is lifted,
these two functions can only differ by a phase factor that does not 
depend on $\mb r$,
\be
\Psi^{(g)}_{g^{-1}\mb k}(\mb r) = \exp[i \alpha_g(\mb k)]
\Psi_{g^{-1}\mb k}(\mb r).
\ee
This  argument also applies to time reversal operation,
\begin{eqnarray}\label{timePsi}
{\cal K}: \Psi_{\mb k}(\mb r) &=& (-i \sigma_y) 
U^*_{\mb k}(\mb r) e^{-i \mb k \mb r}
\equiv \Psi^{({\cal K})}_{-\mb k}(\mb r).
\end{eqnarray}
Let us define $t({\mb k})$ as the phase factor in
\be
\Psi^{({\cal K})}_{-\mb k}(\mb r) = t({\mb k}) \Psi_{-\mb k}(\mb r),
\qquad t^*({\mb k}) = t^{-1}({\mb k}).
\ee
Using the antilinear property of the time reversal operator 
${\cal K}: a \Psi_{\mb k}(\mb r) = a^* {\cal K}: 
\Psi_{\mb k}(\mb r)$ and the fermionic nature of $\Psi_{\mb k}(\mb r)$, 
one obtains,
\begin{eqnarray}
{\cal K}^2: \Psi_{\mb k}(\mb r) &=& - \Psi_{\mb k}(\mb r)\nn\\
 & = & t^*({\mb k}) \,\, {\cal K}: \Psi_{-\mb k}(\mb r) \nn\\
&=& t^*({\mb k}) t({-\mb k}) \Psi_{\mb k}(\mb r)
\end{eqnarray}
Thus, $t({\mb k})$ is an odd function of $\mb k$,
\be\label{oddt}
t(-\mb k) =  - t(\mb k)
\ee

Before discussing the interaction part of the Hamiltonian, we formulate the 
symmetry properties of the Hamiltonian (\ref{Ham1}) using the language of 
second quantization [See, e.g., Refs.~\onlinecite{Landau77,Ambegaokar69}].
The $N$-particle wave 
function $\Phi_N$ is the anti-symmetrized product of $N$ single-particle 
functions $\Psi_{\mb k} (\mb r)$ taken at $N$ different points $\mb k$. 
By writing $(\ldots, 1_{\mb k},\ldots)$ as the argument of $\Phi_N$ we 
emphasize that it describes a state in which a single-particle state with
quantum number $\mb k$ is filled. By definition,
the creation operator $c^\dagger_{\mb k}$,
\be\label{dagger_def}
c^\dagger_{\mb k} \Phi_{N-1}(\ldots, 0_{\mb k},\ldots) = \pm
\Phi_N(\ldots, 1_{\mb k},\ldots),
\ee
where the sign is defined by the antisymmetrization. In particular,
we note that the matrix elements of the creation operator are real numbers.
Using the definition~(\ref{dagger_def}) and the transformation 
properties of 
$\Psi_{\mb k}(\mb r)$, we obtain\cite{Messiah62}
\begin{eqnarray}
g: c^\dagger_{\mb k} &=& \exp[i \alpha_g(\mb k)]\, c^\dagger_{g^{-1} \mb k},
\nn\\
g: c_{\mb k} &=& \exp[-i \alpha_g(\mb k)]\, c_{g^{-1} \mb k},
\nn\\
{\cal K}: c^\dagger_{\mb k}  &=& t(\mb k) c^\dagger_{-\mb k} 
\equiv c^{\dagger({\cal K})}_{-\mb k},\nn\\
{\cal K}: c_{\mb k}  &=& t^*(\mb k) c_{-\mb k},\nn\\
U_1(\theta): c^\dagger_{\mb k} &=& e^{-i\theta} c^\dagger_{\mb k},\nn\\
U_1(\theta): c_{\mb k} &=& e^{i\theta} c_{\mb k},\label{trans_op}
\end{eqnarray}
where $U_1(\theta)$ is a gauge transformation.
The transformation properties of $c_{\mb k}$ follow from the fact that 
the number operator $c^\dagger_{\mb k} c_{\mb k}$ must be a scalar.

The one-band single-particle Hamiltonian can be expressed as
\be
{\cal H}_1 = \sum_{\mb k} \xi_{\mb k} c^\dagger_{\mb k}c_{\mb k},
\ee
where $\xi_{\mb k} = \eps_{\mb k} - \mu$, $\eps_{\mb k}$ is the eigenvalue of
the 
Hamiltonian~(\ref{Ham1}) corresponding to momentum $\mb k$,  and $\mu$ is the 
chemical potential. 
We note that $c^\dagger_{\mb k}$ and $c_{\mb k}$ are both two 
component spinors,  and have all the general properties of spin-$\frac 1 2$ 
fermion operators, even though additional quantum numbers (such as
spin) are omitted since we consider only one band.

\subsection{Pairing term}
If the interband interaction is neglected, the interaction term is
\be\label{HamPair}
{\cal H}_\text{pair} = \frac 1 2 \sum_{\mb k_1, \mb k_2} V_2(\mb k_1, \mb k_2)
c^\dagger_{-\mb k_1}c^\dagger_{\mb k_1}c_{\mb k_2} c_{-\mb k_2},
\ee
where $V_2(\mb k_1, \mb k_2) = \langle \Psi_{-\mb k_1} \Psi_{\mb k_1}|
\widehat V_2|\Psi_{-\mb k_2} \Psi_{\mb k_2} \rangle$ 
is the two-particle matrix element. 
In the weak-coupling approach,\cite{Sigrist91} one introduces the mean field
potential
\be
\Delta(\mb k) = - \sum_{\mb k'} V_2(\mb k, \mb k') 
\langle c_{\mb k'} c_{-\mb k'} \rangle
\ee 
and the interaction (\ref{HamPair}) is approximated by
\be\label{Ham2}
{\cal H}_2 = \frac 1 2 \sum_{\mb k} [\Delta(\mb k) c^\dagger_{\mb k}
c^\dagger_{-\mb k} + \Delta^*(\mb k) c_{-\mb k} c_{\mb k}],
\ee
where a constant term is neglected.
It follows immediately from the anticommutation of the fermion operators
that
\be\label{oddDel}
\Delta(-\mb k) = -\Delta(\mb k)
\ee

One can derive the other transformation properties of $\Delta(\mb k)$ from
the fact that the full Hamiltonian ${\cal H}={\cal H}_1+{\cal H}_2$ is 
invariant under space group operations, time-reversal and gauge 
transformations. Using~(\ref{trans_op}), one obtains
\be\label{transDg}
g: \Delta(\mb k) = \Delta(g^{-1} \mb k) \exp[- i \alpha_g(\mb k) 
- i \alpha_g (-\mb k)]
\ee
It will be seen already from a simple example given in Sec.~\ref{secLim} that 
the phase factor in Eq.~(\ref{transDg}) is not trivial. In general, its 
dependence on $g$ and $\mb k$ cannot
be eliminated. Therefore, the function $\Delta(\mb k)$ \emph{does not} 
transform according to irreducible representations of the space (point) group. 
For crystals with a centre of symmetry, a comprehensive discussion of this 
property has been given by Blount.\cite{Blount85} Instead of the pairing 
potential $\Delta_{\mu\nu}$ which transforms like 
$\langle \psi_\mu, \psi_\nu \rangle$, an auxiliary object which 
transforms like $\langle {\cal K}: \psi_\mu, \psi_\nu \rangle$ is introduced. 
Here $\mu$ and $\nu$ denote the set of single-particle quantum numbers.
Following this idea, we define function $\chi(\mb k)$ by
\be\label{DeltaForm}
\Delta(\mb k) = \chi(\mb k) t(\mb k).
\ee
Then, using the definition of $c^{\dagger({\cal K})}_{-\mb k}$ from the third 
Eq.~(\ref{trans_op}) the first term in ${\cal H}_2$ is expressed as
\be\label{firstT}
\frac 1 2 \sum_{\mb k} \chi(\mb k) t(\mb k) c^\dagger_{\mb k} 
c^\dagger_{-\mb k} = \frac 1 2 \sum_{\mb k} \chi(\mb k) c^\dagger_{\mb k} 
c^{\dagger({\cal K})}_{-\mb k}
\ee
Using the commutation property $g {\cal K} = {\cal K} g$, it is straightforward
to show that\cite{Samokhin04} 
$
g: c^{\dagger({\cal K})}_{-\mb k} = \exp[-i\alpha_g(\mb k)]
c^{\dagger({\cal K})}_{- g^{-1} \mb k}
$. 
Thus, from the invariance of~(\ref{firstT}), we find
\be\label{trans_chi}
g: \chi(\mb k) = \chi(g^{-1} \mb k).
\ee

Eq.~(\ref{trans_chi}) is the basis of the 
group-theoretical classification of SC 
states in noncentrosymmetric crystals. $\chi(\mb k)$ can be expanded in terms 
of basis functions $\chi_i(\mb k)$ 
of irreducible representations of the space (point) group. 
At this point we restrict our consideration to homogeneous SC states for 
simplicity, so that only the point group is involved,
\be
\chi(\mb k) = \sum_i \eta_i \chi_i(\mb k)
\ee
where $\eta_i$ will be identified as the components of the order parameter.
It follows from Eqs.~(\ref{oddt}),~(\ref{oddDel}) and~(\ref{DeltaForm}) that 
$\chi(-\mb k)=\chi(\mb k)$. Examples of even basis functions 
$\chi_i(\mb k)$ for the irreducible representations of point groups $C_{4v}$ 
for \ce\ and $C_2$ for UIr are given in Table~\ref{tbl}.

\begin{table}
\caption{\label{tbl}Even basis functions for irreducible representations (IR)
of point groups $C_{4v}$ and $C_2$. The notations for IR and their character 
tables can be found in Ref.~\onlinecite{Landau77}. $a$, $b$, and $c$ denote 
lattice constants.}
\begin{ruledtabular}
\begin{tabular}{cll}
IR & Non-periodic & Periodic\\
\hline
$C_{4v}$\\
$A_1$ & 1, $k_x^2 + k_y^2$, $k_z^2$ & $\cos k_x a + \cos k_y a$, $\cos k_z c$\\
$A_2$ & $k_xk_y(k_x^2-k_y^2)$ & $\sin k_xa \sin k_ya (\cos k_xa-\cos k_ya)$\\
$B_1$ & $k_x^2 - k_y^2$ & $\cos k_x a - \cos k_y a$\\
$B_2$ & $k_x k_y$ & $\sin k_xa \sin k_ya$\\
$E$ & $\left\{\begin{array}{l} k_x k_z \\ k_y k_z \end{array} \right.$ &
$\left\{\begin{array}{l} \sin k_xa \sin k_zc \\ \sin k_ya \sin k_zc 
\end{array} \right.$\\
\hline
$C_2$\\
$A_1$ & 1, $k_x^2$, $k_y^2$, $k_z^2$, & $\cos k_xa$, $\cos k_yb$, $\cos k_zc$\\
      & $k_xk_y$ & $\sin k_xa \sin k_yb$\\
$A_2$ & $k_x k_z$, $k_y k_z$ & $\sin k_xa \sin k_zc$, $\sin k_yb \sin k_zc$
\end{tabular}
\end{ruledtabular}
\end{table}

The other transformation 
properties of $\chi(\mb k)$ also follow from~(\ref{trans_op}) 
and~(\ref{firstT}),
\be
{\cal K}: \chi(\mb k) = \chi^*(\mb k), \qquad U_1(\theta): \chi(\mb k) = 
e^{2i\theta} \chi(\mb k),
\ee
where we again use antilinearity of ${\cal K}$. As is usually done in 
the 
Ginzburg-Landau approach, the transformation properties of $\chi(\mb k)$ are 
reformulated as transformations of $\eta_i$.\cite{Sigrist91}
If $\chi_i(\mb k)$ are chosen real, then $g:\eta_i = D_{ik} \eta_k$,
${\cal K}: \eta_i =\eta_i^*$ and $U_1(\theta): \eta_i = e^{2i\theta} \eta_i$,
where $D$ is the matrix corresponding to $g$ in the representation chosen.

\subsection{Possible forms of the Ginzburg-Landau potential for \ce}

Even though $C_{4v}$ lacks inversion symmetry, the homogeneous terms in the
Ginzburg-Landau potential (GLP) coincide with those of
$D_{4h}$,\cite{Sigrist91} because the GLP has to be invariant with respect to 
gauge transformations, in particular $U_1(\pi/2): \chi(\mb k) = - \chi(\mb k)$.
For every representation $\Gamma$ listed in Table~\ref{tbl}, the product 
$\Gamma\otimes\Gamma^*$ contains the representation $A_1$. The $z$-component of
the gradient operator
\be
\mb D = \nabla - \frac{2ie}{ c} \mb A
\ee
also transforms according to $A_1$, therefore the GLP density contains a term
linear in $D_z$,
\be
\sum_i \eta_i^* D_z \eta_i + \eta_i D_z^* \eta^*_i.
\ee 
However, this terms is equal to the derivative 
$\frac{\partial}{\partial z} \sum_i |\eta_i|^2$. Hence, after integration over
the sample, its contribution to the GLP vanishes.

For all one-dimensional representations of $C_{4v}$, the GLP is
\begin{eqnarray}
F_{1D} = F_0 + &\int& d\mb r [K_1 (|D_x \eta|^2 + |D_y \eta|^2) 
+ K_2 |D_z \eta|^2\nn\\
&&  + \alpha |\eta|^2 +\beta |\eta|^4].
\end{eqnarray}
For the two-dimensional representation $E$,
\be
\begin{array}{l}
F_{2D} = F_0 + \int d\mb r 
\{K_1 (|D_x \eta_1|^2 + |D_y \eta_2|^2) \\
+ K_2(|D_x \eta_2|^2 + |D_y \eta_1|^2)
+ K_3 [(D_x\eta_1)(D_y\eta_2)^* + \text{c.c.}] \\
+ K_4 [(D_x\eta_2)(D_y\eta_1)^* 
+ \text{c.c.}]
 + K_5(|D_z \eta_1|^2 + |D_z \eta_2|^2) \\
+ \alpha (|\eta_1|^2+|\eta_2|^2) 
+ \beta_1 (|\eta_1|^2+|\eta_2|^2)^2 +\beta_2(\eta_1^*\eta_2-\eta_1\eta_2^*)^2\\
+\beta_3|\eta_1|^2|\eta_2|^2\}.
\end{array}
\ee

\subsection{Green's functions}
The most straightforward way to calculate the normal $G(\mb k, \tau)$ and 
anomalous $F(\mb k, \tau)$, $F^\dagger(\mb k, \tau)$ Green's 
functions within the weak-coupling approach is to use the Gor'kov 
equations.\cite{Gorkov58,Mahan90}
The Green's functions are defined as follows,
\begin{eqnarray}\label{defGreen}
G(\mb k, \tau)&=&-\langle T_\tau\{c_{\mb k}(\tau)c^\dagger_{\mb k}(0)\} 
\rangle \nn\\ 
F(\mb k, \tau)&=&\langle T_\tau\{c_{\mb k}(\tau)c_{-\mb k}(0)\}
\rangle \nn\\
F^\dagger(\mb k, \tau)&=&\langle T_\tau\{c^\dagger_{-\mb k}(\tau)
c^\dagger_{\mb k}(0)\}\rangle,
\end{eqnarray}
where $T_\tau$ is the imaginary time ordering operator. 
The Gor'kov equations for the Fourier transforms are
\begin{eqnarray}
[i\omega_n-\xi_{\mb k}] G(\mb k, \omega_n)+  
\Delta(\mb k) F^\dagger(\mb k, \omega_n) &=&1,\nn\\ 
\left[i\omega_n+\xi_{\mb k}\right]
 F^\dagger(\mb k, \omega_n)+
\Delta^*(\mb k) G(\mb k, \omega_n) &=& 0,
\end{eqnarray}
where $\omega_n=\pi T (2n+1)$  are the Matsubara frequencies for fermions. 
The equations are easily solved by 
\begin{eqnarray}
G(\mb k, \omega_n)&=&{-(i\omega_n+\xi_{\mb k})}/
{(\omega_n^2+\xi_{\mb k}^2 + |\chi(\mb k)|^2)},\nn\\
F^\dagger(\mb k, \omega_n)&=&{\chi^*(\mb k)t^*(\mb k)}/
{(\omega_n^2+\xi_{\mb k}^2 + |\chi(\mb k)|^2)}, \label{solGreen}
\end{eqnarray}
where we have made use of the fact that $|\Delta(\mb k)|^2 = |\chi(\mb k)|^2$.

Gor'kov and Rashba\cite{Gorkov01} considered a model of superconductivity with
split bands due to 
SOC and an isotropic pairing interaction. They showed that the 
thermodynamics of a such a superconductor is equivalent to that of a 
conventional $s$-wave superconductor. Here we have generalized
this result to the anisotropic case. As is evident from the denominator in 
Eq.~(\ref{solGreen}), the thermodynamics of a noncentrosymmetric superconductor
are governed by the gap $|\chi(\mb k)|$. 

\section{\label{secJos}Josephson effect}
We consider tunneling of SC electrons between a superconductor with 
non-degenerate bands and a light conventional superconducting metal 
like Nb, in which electronic bands are spin degenerate.
The tunneling Hamiltonian is
\be
{\cal H}_T = \sum_{\mb k_1, \mb k_2, s} [T_s(\mb k_1, \mb k_2)
c^\dagger_{\mb k_1} a_{\mb k_2, s} + T_s^*(\mb k_1, \mb k_2)
a^\dagger_{\mb k_2, s}c_{\mb k_1}],
\ee
where, $a^\dagger_{\mb k, s}$ creates an electron with spin $s$ in the 
conventional superconductor and $T_s(\mb k_1, \mb k_2)$ is the 
tunneling matrix element. By applying the time reversal operation to the
first term in ${\cal H}_T$, we obtain
\be
\begin{array}{l}
\displaystyle
\sum_{\mb k_1, \mb k_2, s, s'} T^*_s(\mb k_1, \mb k_2)
t(\mb k_1) c^\dagger_{-\mb k_1} (-i\sigma_y)_{s s'} a_{-\mb k_2, s'}\\
\displaystyle
= \sum_{\mb k_1, \mb k_2, s, s'} t(-\mb k_1) (i\sigma_y)_{s s'} 
T^*_{s'}(-\mb k_1, -\mb k_2) c^\dagger_{\mb k_1} a_{\mb k_2, s}\\
\end{array}
\ee
Therefore,
\be\label{Tprop}
T_s (-\mb k_1, -\mb k_2) = t(\mb k_1) \sum_{s'} (i \sigma_y)_{ss'} 
T^*_{s'} (\mb k_1, \mb k_2).
\ee

The following derivation follows that for conventional superconductors 
given in Refs.~\onlinecite{Sigrist91},~\onlinecite{Mahan90}, which is in turn 
based on the
formalism originally proposed by Ambegaokar and Baratoff.\cite{Ambegaokar63}
For simplicity, we restrict our consideration to the case of zero voltage.
The current flowing between the two superconductors is 
$j=e \langle \partial N(t)/\partial t \rangle = 
e i \langle[{\cal H}_T, N]\rangle$, where $N(t) = e^{i {\cal H}_T t} 
\sum_{\mb k}c^\dagger_{\mb k}c_{\mb k} e^{-i {\cal H}_T t}$ is the 
time-dependent particle number operator. 

Following the same steps as described in Ref.~\onlinecite{Mahan90}, we obtain
\begin{widetext}
\be
j=2 e T \text{Im} \sum_{\mb k_1, \mb k_2,s,s',n} T_s(\mb k_1, \mb k_2) 
T_{s'}(-\mb k_1, -\mb k_2) F^{\dagger}(\mb k_1, \omega_n)
F^{(c)}_{s s'} (\mb k_2, \omega_n),
\ee
where 
$
F^{(c)}_{s s'} (\mb k, \omega_n) = \psi(\mb k)(i\sigma)_{s s'}/(\omega_n^2 + 
\epsilon^2_{\mb k} +|\psi(\mb k)|^2)
$
is the anomalous Green's function for a conventional superconductor with
the single-particle spectrum $\epsilon_{\mb k}$ and singlet gap
function $\psi(\bf k)$.\cite{Sigrist91}
Using~(\ref{solGreen}) and~(\ref{Tprop}), we obtain
\be\label{Josepshon}
j= - 2 e T \text{Im} \sum_{\mb k_1, \mb k_2} \sum_n
\frac{\chi^*(\mb k_1)\psi(\mb k_2)}
{(\omega_n^2+\xi_{\mb k_1}^2 +|\chi(\mb k_1)|^2) (\omega_n^2 + 
\epsilon^2_{\mb k_2} +|\psi(\mb k_2)|^2) } 
\sum_s |T_s(\mb k_1, \mb k_2)|^2.
\ee
\end{widetext}
The sum over Matsubara frequencies can be easily calculated if needed by using 
the formula $ 2 T \sum_n (\omega_n^2+E^2)^{-1}= \tanh (E/2 T)/E$.
Therefore, the final expression for the Josephson current does not depend on
$t(\mb k)$. Once again, we see that 
the superconductor with SOC split bands behaves as a 
singlet superconductor with gap function $\chi(\mb k)$.

The tunneling Hamiltonian ${\cal H}_T$ is invariant with respect to the
operations $g$ of  the point group which leave the surface
invariant. Thus,
\be
g: T_s(\mb k_1, \mb k_2) = e^{-i \alpha_g(\mb k_1)} {\cal D}(g)_{s s'} 
T_{s'}(g^{-1}\mb k_1, g^{-1}\mb k_2)
\ee
It follows that 
\be
g: \sum_s |T_s(\mb k_1, \mb k_2)|^2 = 
\sum_s |T_s(g^{-1}\mb k_1, g^{-1}\mb k_2)|^2
\ee
Therefore, $j$ is invariant under the operations $g$ as expected.

\section{\label{secLim}Limit of small spin-orbit coupling}

In order to verify
the general results given in Sec. II, 
it is important to show that they remain valid for crystals with small SOC,
{\em i.e.} for almost degenerate electronic bands. In this Section we consider 
the Rashba
approximation,\cite{Rashba60} because (i) the one particle Hamiltonian is 
then exactly solvable and (ii) we can easily compare our results for 
superconductivity with those known from the literature, where the Rashba 
approximation was used.\cite{Gorkov01,Frigeri04}

In the Rashba approximation, one assumes that the solutions of the Hamiltonian
$H_0=\mb p^2/2m + V(\mb r)$ are plain waves with definite projections of spin
and energy spectrum $\eps_{\mb k}^0$,
and that $\nabla V(\mb r)$ can be replaced by a constant vector directed along
the axis of symmetry $\mb n=(001)$. Therefore Eq.~(\ref{Ham1}) is approximated 
by
\be
H'_1=
\begin{pmatrix}
\eps_{\mb k}^0 & -\beta(k_y +i k_x) \\
-\beta(k_y - i k_x) & \eps_{\mb k}^0
\end{pmatrix},
\ee
where $\beta$ is a constant which characterizes the strength of the SOC. The 
Hamiltonian is diagonalized by the spinors\cite{Gorkov01}
\be
\Psi_{\mb k,\pm} = \frac 1 {\sqrt{2 V}}
\begin{pmatrix}
1 \\ \pm i \exp(i \varphi_{\mb k})
\end{pmatrix}
e^{i \mb k \mb r},
\ee
corresponding to the eigenvalues $\eps_{\mb k}^\pm = \eps_{\mb k}^0 \pm \beta 
|\mb k_\perp|$. Here, $\exp(i \varphi_{\mb k})= (k_x + i k_y)/|\mb k_\perp|$
and $|\mb k_\perp| = \sqrt{k_x^2 + k_y^2}$. In the following, we shall 
consider coupling in `+' band as an example.

In order to illustrate the complicated transformation 
properties~(\ref{transDg}) of $\Delta(\mb k)$, let us consider  as an example
$g = m_x$, the mirror plane perpendicular 
to $x$-axis. Acting by $m_x$ on $\Psi_{\mb k +}$, we obtain
\begin{eqnarray}
\frac 1 {\sqrt{2 V}}
\begin{pmatrix}
0 & -i \\ -i & 0
\end{pmatrix}
\begin{pmatrix}
1 \\ i e^{i \varphi_{\mb k}}
\end{pmatrix}
e^{i (\mb k, m_x \mb r)} \nn\\ =
\frac {e^{i \varphi_{\mb k}}} {\sqrt{2 V}}
\begin{pmatrix}
1 \\ i e^{i \varphi_{m_x \mb k}}
\end{pmatrix}
e^{i (m_x \mb k, \mb r)} & =&   e^{i \varphi_{\mb k}} \Psi_{m_x \mb k, +}
\end{eqnarray}
where $\exp(i \varphi_{m_x \mb k}) = - \exp(-i \varphi_{\mb k})$ was used. 
Hence, the phase factor defined by  Eq.~(\ref{transDg}) is equal to 
$(i k_x+k_y)^2/\mb k^2_\perp$.

By applying time reversal to  $\Psi_{\mb k,+}$ using 
Eq.~(\ref{timePsi}), we find 
\be\label{t_p}
t(\mb k) = i\exp(-i \varphi_{\mb k}).
\ee

The fermion operators corresponding to initial spin-up and spin-down 
states can be expressed in terms of the new band operators as
\be\label{old_op}
c^\dagger_{\mb k \uparrow} = \frac 1 {\sqrt{2}} 
(c^\dagger_{\mb k +}+c^\dagger_{\mb k -}),
\qquad
c^\dagger_{\mb k \downarrow} = - \frac {i e^{-i\varphi_{\mb k}}} {\sqrt{2}} 
(c^\dagger_{\mb k +}-c^\dagger_{\mb k -}).
\ee
At this point we assume that before SOC has been turned on, the electrons 
were paired in the singlet state with gap function $\psi(\mb k)$,
\be
{\cal H}'_2 = \frac 1 2 \sum_{\mb k} \psi (\mb k) c^\dagger_{\mb k \uparrow}
c^\dagger_{-\mb k \downarrow} + \text{h.c.}
\ee
Using~(\ref{old_op}) and~(\ref{t_p}), we obtain
\be\label{pair_p}
{\cal H}'_2 = \frac 1 4 \sum_{\mb k} t(\mb k)\psi (\mb k) 
c^\dagger_{\mb k +} c^\dagger_{-\mb k +} + \text{h.c.} + \ldots,
\ee
where the rest of the terms, which describe pairing in the other band and 
inter-band pairing, are neglected. By comparing~(\ref{pair_p}) 
with~(\ref{Ham2}) we obtain $\Delta(\mb k)$ in the form~(\ref{DeltaForm}) with
$\chi(\mb k) = \psi(\mb k)/2$.

Interestingly, a similar result is obtained if one starts with triplet 
pairing for spin states,\cite{Sigrist91}
\begin{widetext}
\begin{eqnarray}
\displaystyle
{\cal H}'_2 &=& \frac 1 2 \sum_{\mb k} \{[-d_x(\mb k) + i d_y(\mb k)] 
c^\dagger_{\mb k \uparrow} c^\dagger_{-\mb k \uparrow} + 
[d_x(\mb k) + i d_y(\mb k)] c^\dagger_{\mb k \downarrow} c^\dagger_{-\mb k 
\downarrow} 
+ d_z(\mb k) (c^\dagger_{\mb k \uparrow} c^\dagger_{-\mb k \downarrow} +
c^\dagger_{\mb k \downarrow} c^\dagger_{-\mb k \uparrow})\} + \text{h.c.}
\nn\\
&=& \frac 1 2 \sum_{\mb k} t(\mb k) [d_y(\mb k) \hat k_x - d_x (\mb k) \hat
k_y] c^\dagger_{\mb k +} c^\dagger_{-\mb k +} + \text{h.c.} + \ldots,
\end{eqnarray}
\end{widetext}
where $\hat k_i = k_i/|\mb k_\perp|$ and again, only  terms corresponding
coupling within the `$+$' band are kept. Therefore, we again find $\Delta(\mb k)$ 
in the form~(\ref{DeltaForm}) with 
\be
\chi(\mb k) = d_y(\mb k) \hat k_x - 
d_x (\mb k) \hat k_y.
\label{pizda}
\ee 

Frigeri \emph{et al.}\cite{Frigeri04} used weak coupling theory to
show that a phase transition to 
the triplet pairing state $\mb d(\mb k)= (- k_y, k_x,0)$ 
(which transforms according to the representation $A_2$)
in \ce\ is not 
suppressed by small but finite $\beta$.  According to
(\ref{pizda}), this corresponds to 
$\chi(\mb k) \propto \hat k_x^2 + \hat k_y^2$, which is isotropic and fully 
gapped and transforms like $A_1$.
By using the
vectors $\mb d(\mb k)$ corresponding to the rest of the irreducible
representations of $C_{4v}$ (which are similar to those 
for $D_{4h}$ and are given in
Ref. \onlinecite{Sigrist91}), the corresponding functions
$\chi(\mb k)$ may be obtained from Eq.~(\ref{pizda}) and are
listed  in Table~\ref{tbl2}. 
Therefore, in general, 
starting from $\mb d(\mb k)$ corresponding to a representation $\Gamma$, 
we generate $\chi(\mb k)$ using Eq.~(\ref{pizda})
corresponding to another representation
$\Gamma'$, while in the singlet case, $\chi(\mb k)$ and $ \psi(\mb k)/2$ 
belong to the same representation because they are proportional. 
The difference is due to the transformation
properties of the paired state under spin rotation, 
which was used to diagonalize the Hamiltonian $H_1'$: 
the singlet state transforms into itself as a scalar, whereas the 
triplet state transforms as a spin-1 state.\cite{DiffNote}

\begin{table}[b]
\caption{\label{tbl2}Basis functions for triplet superconductivity in $C_{4v}$
crystals with non-degenerate bands and 
vanishing SOC and the corresponding even basis functions 
determined by Eq.~(\ref{pizda}). Note that $\Delta(\bf k)$ is defined by
Eq.~(\ref{DeltaForm}) even for the triplet case. }
\begin{ruledtabular}
\begin{tabular}{cllc}
$A_1$ & $\mb d (\mb k) = (k_x,k_y,0)$ & --\\
      & $\mb d (\mb k) = (k_x^3,k_y^3,0)$ & 
                $\chi(\mb k) = k_x k_y (k_x^2-k_y^2)$ & $A_2$ \\
$A_2$ & $\mb d (\mb k) = (-k_y,k_x,0)$ & $\chi(\mb k) = k_x^2+k_y^2$ & $A_1$\\
$B_1$ & $\mb d (\mb k) = (k_x, - k_y,0)$ & $\chi(\mb k) = k_x k_y$ & $B_2$ \\
$B_2$ & $\mb d (\mb k) = (k_y, k_x,0)$ & $\chi(\mb k) =k_x^2 - k_y^2$ & $B_1$\\
$E$ & $\mb d_1(\mb k) = (k_z,0,0)$ & $\chi_1(\mb k) = k_y k_z$ & $E$\\
 & $\mb d_2(\mb k) = (0,k_z,0)$ & $\chi_2(\mb k) = k_x k_z$
\end{tabular}
\end{ruledtabular}
\end{table}

\section{\label{secDisc}Discussion}

A close analogy between the superconductors under consideration and usual 
centrosymmetric singlet superconductors should not be surprising. A particle 
described by the wave function~(\ref{Psik}) which is a solution of 
Hamiltonian~(\ref{Ham1}) has spin
$\mb s (\mb k) = \frac 1 2 \langle \Psi_{\mb k}| \boldsymbol \sigma 
|\Psi_{\mb k}\rangle$. Since $\Psi_{-\mb k}$ and $\Psi_{-\mb k}^{({\cal K})}$ 
differ only by a constant phase factor, we have
\begin{eqnarray}
\mb s(-\mb k) &=& \frac 1 2 \langle \Psi_{-\mb k}| \boldsymbol \sigma 
|\Psi_{-\mb k}\rangle = 
\frac 1 2 \langle \Psi_{-\mb k}^{({\cal K})}| \boldsymbol \sigma 
|\Psi_{-\mb k}^{({\cal K})}\rangle \nn\\
&=& {\cal K}: \frac 1 2 \langle \Psi_{\mb k}| \boldsymbol \sigma^{({\cal K})} 
|\Psi_{\mb k}\rangle = \frac 1 2 \langle \Psi_{\mb k}| \boldsymbol 
\sigma^{({\cal K})} |\Psi_{\mb k}\rangle^*\nn\\
& = & - \frac 1 2 \langle \Psi_{\mb k}| \boldsymbol \sigma |\Psi_{\mb k}\rangle
 = -\mb s (\mb k),
\end{eqnarray}
where we have 
used the fact that diagonal matrix elements of the spin operator are 
real and $\boldsymbol \sigma^{({\cal K})} = (-i \sigma_y)\boldsymbol \sigma^*
(-i \sigma_y)^{-1} = - \boldsymbol \sigma$. Therefore, particles with 
opposite momenta within the same band always have opposite 
spin [See also Ref.~\onlinecite{Saxena04}].

It was first noted by Anderson\cite{Anderson59} that for crystals in which 
either the momentum 
$\mb k$ or spin is not a good quantum number, as in dirty superconductors, a
one particle state should be paired with its time reversal. Using this idea,
Samokhin \emph{et al.}\cite{Samokhin04} define $c^\dagger_{-\mb k}$ as time
reversal of $c^\dagger_{\mb k}$.  This in turn leads to the conclusion that 
$\Delta(\mb k)$ is an odd function which 
transforms according to the irreducible 
representations of the point group.  Then, the gaps in the quasiparticle 
spectrum of the  superconducting states described by all one dimensional 
representations of $C_{4v}$ have line nodes. This, in particular, rules out
the possibility of an isotropic full gap, and therefore
contradicts the theoretical results of 
Refs.~\onlinecite{Gorkov01,Frigeri04} and the experimental
results for Cd$_2$Re$_2$O$_7$. 
The origin of the discrepancy is that Anderson's statement is correct up to a 
phase factor $t(\mb k)$, which is not important in the normal state due to 
gauge invariance but cannot be ignored in the superconducting 
state.\cite{AndersonNote} In other words, one cannot define $c^\dagger_{\mb k}$
in an asymmetric unit of the Brillouin zone and then use the symmetry 
elements (including time reversal) to define the states in the rest of the 
Brillouin zone. Such a procedure would lead to a discontinuity of the phase of
the wave function on the boundaries of the asymmetric units. Also, if 
$c^\dagger_{\mb k}$ is paired with ${\cal K} c^\dagger_{\mb k}$, then 
${\cal K} c^\dagger_{\mb k}$ must be paired with 
${\cal K}^2 c^\dagger_{\mb k}=- c^\dagger_{\mb k}$, 
which is a contradiction. Instead, if 
$c^\dagger_{\mb k}$ describes the particle with momentum $\mb k$ and 
$\mb k' = g^{-1} \mb k$, where $g$ is a point group element or time reversal, 
then the particle with momentum $\mb k'$ must be described by 
$c^\dagger_{\mb k'} = c^\dagger_{g^{-1}\mb k}$, which is proportional but not 
equal to $g: c^\dagger_{\mb k}$.

The formalism developed in this paper can be directly applied to 
Cd$_2$Re$_2$O$_7$, which shows no sign of magnetic ordering and therefore
possesses time reversal symmetry.\cite{Vyaselev02} The low temperature 
structure has symmetry $I4_122$,\cite{Yamaura02} with point group $D_4$, which 
is isomorphic to $C_{4v}$. The superconducting order parameter corresponds
to the 
representation $A_1$ since no nodes in the quasi-particle spectrum have been
found.

In the application to CePt$_3$Si, this theory should be slightly modified to
include the effect of antiferromagnetic ordering ($T_N = 2$ K).\cite{Bauer04} 
A neutron scattering study\cite{Metoki04} revealed that the antiferromagnetic
structure is characterized by the wave vector $\mb Q = (0,0,\pi/c)$. Therefore,
the normal state just above $T_c$ is invariant under the operation 
$\tau(\mb c){\cal K}$ instead of ${\cal K}$, where $\tau(\mb c)$ is a
lattice translation along $z$-axis. This leaves valid all of the results of our 
discussion, since we consider translationally invariant 
superconducting states, and for the antiferromagnetic case, one still has 
$\eps_{\mb k}=\eps_{-\mb k}$. However, the presence of antiferromagnetic order
can provide certain clues about possible superconducting state. It has been 
shown for zero SOC that the singlet superconducting state with gap function
$\psi(\mb k)$ is energetically favoured if $\psi (\mb k + \mb Q) = 
- \psi (\mb k)$ [See Ref.~\onlinecite{Sigrist91} and references therein].
Hence, the periodic basis functions in Table I, which depend on $k_z$, 
may be good candidates. 

Even more exciting is the situation realized in UIr.\cite{Akazawa04}
At ambient pressure it is ferromagnetic with Curie temperature of 46 K.
Superconductivity occurs at high pressures of 2.6-2.7 GPa. One possibility
is that the phase transition lines from the normal paramagnetic to 
ferromagnetic
state and from the normal paramagnetic to superconducting state meet at a quantum
critical point. In that case superconducting states can again be classified 
with respect to the symmetry of time reversal invariant paramagnetic state.
On the other hand, ferromagnetism can survive high pressures and coexist with
the superconducting state. Then, time reversal 
symmetry is completely broken. Since the centre of symmetry is also missing,
in general $\eps_{\mb k} \neq \eps_{-\mb k}$. Therefore, a zero-field analog 
of the inhomogeneous Larkin-Ovchinnikov-Fulde-Ferrel state,\cite{Casa04} 
may be realized in this material. On the other hand, 
superconductivity exists in a relatively narrow pressure range. 
Hence, an alternative scenario may imply that $\eps_{\mb k}$ and 
$\eps_{-\mb k}$ are \emph{accidentally} degenerate for a part of the Fermi
surface. In either case, superconductivity in UIr deserves further 
investigation.

In conclusion, we have considered the symmetry properties of the gap
function in superconductors with lifted spin degeneracy.
We have shown that phase factors which appear with 
the gap function $\Delta(\mb k)$ under point group operations and time reversal
may be handled by the introduction of an even auxiliary function 
$\chi(\mb k)$, which transforms according to the irreducible representations
of the point group. 
This function defines the behaviour of the superconductor in terms of
thermodynamic and tunneling properties. 

\emph{Note added.} The criticism of Ref.~\onlinecite{Samokhin04} presented
in this Article is addressed in Ref.~\onlinecite{Samokhin04E}.

\begin{acknowledgments}
It is our pleasure to acknowledge very useful discussions with D.F. Agterberg
and D.J. Singh. This work was supported by NSERC Canada. 
\end{acknowledgments}


\begin{thebibliography}{00}
\bibitem{NoteTh} This may not be the case for 
(Y$_{1-x}$Th$_x$)$_2$C$_{3-y}$ and (La$_{1-x}$Th$_x$)$_2$C$_{3-y}$, in
which strong SOC is expected. As reported in 
Refs.~\onlinecite{Krupka69, Giorgi70}, superconductivity with $T_c> 10$ K 
survives the doping level up to $x=0.9$. Unfortunately, to the best of our 
knowledge, band structure calculations have 
not been reported for the Th doped compounds.
\bibitem{AndersonNote} Time reversal of a fermion state $|\phi\rangle$ is 
defined as $|\phi\rangle^*$ in Ref.~\onlinecite{Anderson59}. Strictly speaking,
it must be accompanied by changing the sign of one of the spinor components. 
Only then, the correct property ${\cal K}^2|\phi\rangle = -|\phi\rangle$ is
obtained.
\end{thebibliography}
\end{document}